\documentclass[article]{jss}

\newcommand{\R}{{\cal R}}
\newcommand{\M}{{\cal M}}

\newcommand{\bX}{{\bf X}}
\newcommand{\bZ}{{\bf Z}}
\newcommand{\bM}{{\bf M}}
\newcommand{\bG}{{\bf G}}
\newcommand{\bD}{{\bf D}}
\newcommand{\bJ}{{\bf J}}
\newcommand{\bU}{{\bf U}}
\newcommand{\bV}{{\bf V}}
\newcommand{\bF}{{\bf F}}
\newcommand{\bE}{{\bf E}}

\newcommand{\bx}{{\bf x}}
\newcommand{\by}{{\bf y}}
\newcommand{\bz}{{\bf z}}
\newcommand{\bu}{{\bf u}}

\author{Marie Chavent \\ University of Bordeaux \And 
         Vanessa Kuentz\\CEMAGREF Bordeaux \AND
 Beno\^it Liquet\\ University of Bordeaux\And 
 J\'er\^ome Saracco \\University of Bordeaux}
\title{ \pkg{ClustOfVar}: An \proglang{R} Package for the Clustering of Variables}

\Plainauthor{Marie Chavent,Vanessa Kuentz, Beno\^it Liquet,  J\'er\^ome Saracco} %% comma-separated
\Plaintitle{ ClustOfVar: An  R  Package for the Clustering of Variables} %% without formatting

\Abstract{
Clustering of variables is as a way to arrange variables into homogeneous clusters, i.e., groups of variables which are strongly related to each other and thus bring the same information. These approaches can then be useful for dimension reduction and variable selection.  Several specific methods have been developed for the clustering of numerical variables. However concerning qualitative variables or  mixtures of quantitative and qualitative variables, far fewer methods have been proposed.  The  \proglang{R}  package  \pkg{ClustOfVar} was specifically developed  for this purpose.   The homogeneity criterion of a cluster  is defined as the sum of  correlation ratios (for qualitative variables) and squared correlations (for quantitative variables) to a synthetic quantitative variable, summarizing  ``as good as possible'' the variables in the cluster. This synthetic variable is the first principal component obtained with the PCAMIX method. Two algorithms for the clustering of variables are proposed: iterative relocation algorithm and ascendant  hierarchical clustering.
We also propose a bootstrap approach in order to determine suitable numbers of clusters.
We illustrate the  methodologies   and  the associated package on small datasets.\\

}

\Address{
  Marie Chavent\\
- Univ. Bordeaux, IMB, UMR 5251, F-33400 Talence, France.\\
- CNRS, IMB, UMR 5251, F-33400 Talence, France. \\
- INRIA, F-33400 Talence, France. \\
  E-mail: \email{Marie.Chavent@u-bordeaux2.fr}\\

Vanessa Kuentz\\
Cemagref, UR ADBX, F-33612 Cestas Cedex, France\\
  E-mail: \email{vanessa.kuentz@cemagref.fr}\\

Benoit Liquet\\
- INSERM, ISPED, Centre INSERM U-897-Epidemiologie-Biostatistique, Bordeaux, F-33000\\
- Univ. Bordeaux, ISPED, Centre INSERM U-897-Epidemiologie-Biostatistique, Bordeaux, F-33000, France\\
 E-mail: \email{Benoit.Liquet@isped.u-bordeaux2.fr}\\

Jerome Saracco\\
- IPB, IMB, UMR 5251, F-33400 Talence, France.\\
- CNRS, IMB, UMR 5251, F-33400 Talence, France. \\
- INRIA, F-33400 Talence, France. \\
 E-mail: \email{jerome.saracco@math.u-bordeaux1.fr}
}
\begin{document}

%% include your article here, just as usual
%% Note that you should use the \pkg{}, \proglang{} and \code{} commands.

%%%%%%%%%%%%%%%%%%%%%%%%%%%%%%%%%%%%%%%%%%
\section[Introduction]{Introduction}
%%%%%%%%%%%%%%%%%%%%%%%%%%%%%%%%%%%%%%%%%%

%% Note: If there is markup in \(sub)section, then it has to be escape as above.

Principal Component Analysis (PCA) and Multiple Correspondence Analysis (MCA) are appealing statistical tools for multivariate description of respectively numerical and categorical data. Rotated principal components fulfill the need to get more interpretable components. Clustering of variables is an alternative since it makes possible to arrange variables into homogeneous clusters and thus to obtain meaningful structures. From a general point of view, variable clustering lumps together variables which are strongly related to each other and thus bring the same information. Once the variables are clustered into groups such that attributes in each group reflect the same aspect, the practitioner may be spurred on to select one variable from each group. One may also want to construct a synthetic variable. For instance in the case of quantitative variables, a solution is to realize a PCA  in each cluster and to retain the first principal component as the synthetic variable of the cluster. 

A simple and frequently used approach for clustering a set of variables is to calculate the dissimilarities between these variables  and to apply a classical cluster analysis method to this  dissimilarity matrix. We can cite the functions  \code{hclust} of the  \proglang{R} package \pkg{stats} \citep{R} and \code{agnes} of the package \pkg{cluster}  \citep{cluster} which can be used  for single, complete, average linkage hierarchical clustering. The functions \code{diana} and \code{pam} of the package \pkg{cluster} can also be used for respectively divisive hierarchical clustering and partitioning around medoids  \citep{Kaufman90}. But   the dissimilarity matrix has to be calculated first. For quantitative variables many dissimilarity measures can be used: correlation coefficients (parametric or nonparametric) can be converted to different dissimilarities depending if the aim is to lump together correlated variables regardless of the sign of the correlation or if a negative correlation coeffcient between two variables shows disagreement between them. For categorical variables, many association measures can be used as $\chi^2$, Rand, Belson, Jaccard, Sokal and Jordan among others. Many strategies can then be applied and it can be difficult for the user to choose oneof them. Moreover, no synthetic variable of the clusters are directly provided with this approach.

Besides these classical  methods devoted to the clustering of observations, there exists  methods specifically devoted  to the clustering of variables. The most famous one is the VARCLUS procedure of  the SAS software. Recently specific methods based on PCA  were proposed   by  \citet{Vigneau03} with the name Clustering around Latent Variables (CLV) and  by \citet{Dhillon03} with the name Diametrical Clustering.  But all these specific approaches work only with quantitative data and as far as we know, they are not implemented in \proglang{R}.

The aim of the package \pkg{ClustOfVar} is then to propose  in  \proglang{R}, methods specifically devoted to the clustering of variables with no restriction on the type (quantitative or qualitative) of the variables. The clustering methods developed  in the package work  with a mixture of quantitative and qualitative variables and  also work  for a set exclusively containing  quantitative (or qualitative) variables.  In addition note that missing data are  allowed:  they are replaced by means for quantitative variables and by zeros in the indicator matrix for qualitative variables. 
Two methods are  proposed for the clustering of variables:  a hierarchical clustering algorithm and a k-means type partitioning algorithm are respectively implemented in the functions \code{hclustvar} and  \code{kmeansvar}. These two methods are based on PCAMIX, a principal component method for a mixture of qualitative and quantitative variables \citep{Kiers91}. This method includes the ordinary PCA and MCA as special cases.  Here we use a Singular Value Decomposition (SVD) approach of PCAMIX \citep{Chavent11}.  
These two clustering algorithms aim at maximizing an homogeneity criterion. A cluster of variables is defined as homogeneous when the variables in the cluster are strongly linked to a central quantitative synthetic variable. This link is measured by the squared Pearson correlation for the quantitative variables and by the correlation ratio for the qualitative variables. The quantitative central synthetic variable of a cluster is  the first principal component of PCAMIX applied to all the variables in the cluster. Note that the synthetic variables of the clusters can be used  for dimension reduction or for recoding purpose. Moreover a method based on a bootstrap approach is also proposed to evaluate the stability of the partitions of variables and can be used to determine a suitable number of clusters. It is implemented in the function \code{stability}. 

The rest of this paper is organized as follows. Section~\ref{sec:hom} contains a detailed description of the
homogeneity criterion and a description of the PCAMIX procedure for the determination of the central synthetic variable. Section~\ref{sec:meth} describes the clustering algorithms and the bootstrap procedure. Section~\ref{sec:exple} provides two data-driven examples in order to illustrate the use of the functions and objects of the package \pkg{ClustOfVar}. Finally, section~\ref{sec:conclu} gives concluding remarks.

%%%%%%%%%%%%%%%%%%%%%%%%%%%%%%%%%%%%%%%%%%
\section[The homogeneity criterion]{The homogeneity criterion} \label{sec:hom}
%%%%%%%%%%%%%%%%%%%%%%%%%%%%%%%%%%%%%%%%%%

Let $\{\bx_1,\dots,\bx_{p_1}\}$ be a set of  $p_1$ quantitative variables and $\{\bz_1,\dots,\bz_{p_2}\}$  a set of  $p_2$ qualitative variables.  Let $\bX$ and $\bZ$  be the corresponding  quantitative and qualitative data matrices of dimensions $n \times p_1$ and $n  \times p_2$, where $n$ is the number of observations. For seek of simplicity, we denote $\bx_j \in \R^n$ the $j$-th column of $\bX$ and  we denote $\bz_j \in \M_1\times \ldots \times \M_{p_2}$ the $j$-th column of $\bZ$ with $\M_j$ the set of categories of $\bz_j$.  Let $P_K=(C_1,\dots,C_K)$  be a partition into $K$ clusters of the $p=p_1+p_2$ variables. 

\paragraph{Synthetic variable of a cluster $C_k$.} It is defined as the quantitative variable $\by_k \in \R^n$ the ``most linked'' to all the variables in $C_k$:  
$$
\by_k=\displaystyle \arg\max_{\bu \in \R^n} \left\{\sum_{\bx_j \in C_k} r^2_{\bu,\bx_j} + \sum_{\bz_j \in C_k} \eta^2_{\bu|\bz_j}  \right\},
$$
where $r^2$ denotes the squared Pearson correlation and $\eta^2$ denotes the correlation ratio. More precisely, the correlation ratio $\eta^2_{\bu|\bz_j} \in [0,1]$ measures the part of the variance of $\bu$ explained by the categories of $\bz_j$: 
 $$\eta^2_{\bu|\bz_j}= \frac{\sum_{s \in \M_j} n_s ( {\bar \bu}_{s}- {\bar \bu})^2}{\sum_{i=1}^n (u_i-{\bar \bu})^2},$$
where $n_s$ is the frequency of category $s$, $\bar \bu_{s}$ is the mean value of $\bu$ calculated on the observations belonging to category $s$
 and $\bar \bu$ is the mean of $\bu$. 

We have the following important results (\citet{Escofier79}, \citet{Saporta90}, \citet{Pages04}):
\begin{itemize}
\item $\by_k$ is the first principal component of  PCAMIX applied to  $\bX_k$ and $\bZ_k$, the matrices made up of the columns of $\bX$ and $\bZ$ corresponding to the variables in $C_k$;
\item the empirical variance of $\by_k$ is equal to: $\displaystyle \VAR(\by_k)=\sum_{\bx_j \in C_k} r^2_{\bx_j,\by_k} + \sum_{\bz_j \in C_k} \eta^2_{\by_k|\bz_j}$.
\end{itemize}

The determination of $\by_k$ using PCAMIX is carried on according to the following steps:
\begin{enumerate}
\item Recoding of $\bX_k$ and $\bZ_k$: 
\begin{enumerate}
\item $\tilde{\bX}_k$ is the standardized version of the quantitative matrix $\bX_k$, 
\item $\tilde{\bZ}_k=\bJ\bG\bD^{-1/2}$ is the standardized version of the indicator matrix $\bG$ of the qualitative matrix $\bZ_k$, where  $\bD$ is the diagonal matrix of frequencies of the categories. $\bJ= {\bf I} - \mathbf{1}^\prime \mathbf{1}/n$ is the centering operator where ${\bf I}$ denotes the identity matrix and $\mathbf{1}$ the vector with unit entries. 
\end{enumerate}
\item Concatenation of the two recoded matrices: $\bM_k=(\tilde{\bX}_k|\tilde{\bZ}_k)$.
\item Singular Value Decomposition of $\bM_k$: $\bM_k=  \bU \Lambda \bV'$.
\item  Extraction/calculus of useful outputs:
\begin{itemize}
\item   $\sqrt{n} \bU \Lambda$ is the matrix of the principal component scores of PCAMIX;
\item  $\by_k$ is the first column  $\sqrt{n} \bU \Lambda$;
\item $\displaystyle \VAR(\by_k)=\lambda_1^k$ where $\lambda_1^k$ is the first eigenvalue in $ \Lambda$.
\end{itemize}
\end{enumerate}

Note that we recently developed an   \proglang{R} package named \pkg{PCAmixdata} with a function \code{PCAmix} which provide  the principal components of PCAMIX and a function \code{PCArot} which provide the principal component after rotation.

\paragraph{Homogeneity $H$ of a cluster $C_k$.} 
It is a measure of adequacy between the variables in the cluster  and its central synthetic quantitative variable $\by_k$:
\begin{equation}
H(C_k)=\sum_{\bx_j \in C_k} r^2_{\bx_j,\by_k} + \sum_{\bz_j \in C_k} \eta^2_{\by_k|\bz_j}=\lambda_1^k.
\label{critC}
\end{equation}
Note that the first term (based on the squared Pearson correlation $r^2$) measures the link between  the quantitative variables in $C_k$ and $\by_k$ independently of the sign of the relationship. The second one (based  on the correlation ratio $\eta^2$) measures the link between the qualitative variables in $C_k$  and $\by_k$. 
The homogeneity of a cluster is maximum when all the quantitative variables are correlated (or anti-correlated) to $\by_k$ and when all the correlation ratios of the qualitative variables are equal to 1. It means that all the variables in the cluster $C_k$ bring the same information.

\paragraph{Homogeneity ${\cal{H}}$ of a partition $P_K$.} It is defined as the sum of the homogeneities of its clusters:
\begin{equation}
{\cal{H}}(P_K)=\sum_{k=1}^K H(C_k)=\lambda_1^1+\ldots+\lambda_1^K,
\label{critP}
\end{equation}
where $\lambda_1^1,\ldots,\lambda_1^K$ are the first eigenvalues of PCAMIX applied to the $K$ clusters $C_k$ of $P_K$.

%%%%%%%%%%%%%%%%%%%%%%%%%%%%%%%%%%%%%%%%%%%
\section[The clustering algorithms]{The clustering algorithms}  \label{sec:meth}
%%%%%%%%%%%%%%%%%%%%%%%%%%%%%%%%%%%%%%%%%%%

The aim is  to find a partition of a set of quantitative and/or qualitative variables  such that the variables within a cluster are strongly related to each other. In other words the objective is to find a partition $P_K$ which maximizes the homogeneity function ${\cal{H}}$ defined in (\ref{critP}). For this, a hierarchical and a partitioning clustering algorithms are proposed in the package \pkg{ClustOfVar}. A bootstrap procedure is also proposed to evaluate the stability of the partitions into $K=2, 3, \dots,p-1$ clusters and then to help the user to determine a suitable number of clusters of variables.

\paragraph{The hierarchical clustering algorithm.}
This algorithm builds a set of $p$ nested partitions of variables in the following way:
\begin{enumerate}
\item Step $l=0$: initialization.  Start with the partition in $p$ clusters.
\item Step $l=1,\dots,p-2$: aggregate two clusters of the partition  in $p-l+1$ clusters to get a new partition in $p-l$ clusters. 
For this, choose  clusters  $A$ and $B$ with the smallest dissimilarity $d$ defined as:
\begin{eqnarray}
d(A,B)=H(A)+H(B)-H(A\cup B) =\lambda_A^1+\lambda_B^1-\lambda_{A\cup B}^1.  
\label{aggM}
\end{eqnarray}
This dissimilarity measures the lost of homogeneity observed when the two clusters $A$ and $B$ are merged. 
Using this aggregation measure  the new  partition in $p-l$ clusters maximizes ${\cal{H}}$ among all the partitions in $p-l$ clusters obtained by aggregation of two clusters of the partition in $p-l+1$ clusters.
\item Step $l=p-1$: stop. The partition in one cluster is obtained.
\end{enumerate}
This algorithm is  implemented in the function \code{hclustvar} which builds a hierarchy of the $p$  variables. The function \code{plot.hclustvar} gives the dendrogram of this hierarchy. 
The height of a cluster $C=A\cup B$  in this dendrogram is  defined as $h(C)=d(A,B)$. It is easy to verify that  $h(C) \geq 0$ but the property ``$A \subset B  \Rightarrow h(A) \leq h(B)$'' has not been proved yet. Nevertheless, inversions in the dendrogram have  never been observed in practice neither on simulated data nor on real data sets. Finally the function \code{cutreevar}  cuts this dendrogram and gives one of the $p$ nested partitions according to the number $K$ of cluster  given in input by the user.

\paragraph{The partitioning algorithm.}
This partitioning algorithm requires the definition of a similarity measure between two variables of any type (quantitative or qualitative). We use for this purpose the squared canonical correlation between two data matrices $\bE$ and $\bF$ of dimensions $n \times r_1$ and $n \times r_2$. 
This correlation, denoted by $s$, can be easily calculated as follows:
The procedure for the its determination is simple:
$$s(\bE,\bF)=\left\{\begin{array}{lcl}
 \mbox{first~eigenvalue~of~the~} n\times n   \mbox{~matrix~} \bE\bF^ \prime \bF\bE^\prime & \mbox{~if~} & \min(n,r_1,r_2)=n,\\
 \mbox{first~eigenvalue~of~the~} r_1\times r_1   \mbox{~matrix~} \bE^{\prime}\bF\bF^{\prime}\bE & \mbox{~if~} & \min(n,r_1,r_2)=r_1,\\
 \mbox{first~eigenvalue~of~the~} r_2\times r_2   \mbox{~matrix~} \bF^{\prime}\bE\bE^{\prime}\bF & \mbox{~if~} & \min(n,r_1,r_2)=r_2.
\end{array}\right.$$
This similarity $s$  can  also be defined as follows:
\begin{itemize}
\item[-]  For two quantitative variables $\bx_i$ and $\bx_j$,  let $\bE=\tilde{\bx}_i$ and  $\bF=\tilde{\bx}_j$ where $\tilde{\bx}_i$ and $\tilde{\bx}_j$ are the standardized versions of $\bx_i$ and $\bx_j$. In this case, the squared canonical correlation is the squared Pearson correlation: $s(\bx_i,\bx_j)=r^2_{\bx_i,\bx_j}$.
\item[-]  For one qualitative variable $\bz_i$ and one quantitative variable $\bx_j$, let $\bE=\tilde{\bZ}_i$ and  $\bF=\tilde{\bx}_j$ where $\tilde{\bZ}_i$ is the standardized version of  the indicator matrix $\bG_i$  of the qualitative variable $\bz_i$. 
In this case,  the squared canonical correlation is the correlation ratio: $s(\bz_i,\bx_j)= \eta^2_{\bx_j|\bz_i}$.
\item[-]  For two qualitative variables $\bz_i$ and  $\bz_j$ having $r$ and $s$ categories,  let $\bE=\tilde{\bZ}_i$ and $\bF=\tilde{\bZ}_j$.
In this case,  the squared canonical correlation  $s(\bz_i,\bz_j)$ does not correspond to a well known association measure. Its interpretation is geometrical: the closer to one is $s(\bz_i,\bz_j)$, the closer are the two linear subspaces spanned by the matrices $\bE$ and $\bF$. Then the  two qualitative variables $\bz_i$ and  $\bz_j$ bring similar information.
\end{itemize}
This similarity measure is implemented in the function \code{mixedVarSim}. 

The clustering algorithm implemented in the function \code{kmeansvar} builds  then a partition in $K$ clusters in the following way:
\begin{enumerate}
\item Initialization step: two possibilities are available.
\begin{enumerate}
\item A non random initialization: an initial partition in $K$ clusters is given in input (for instance the partition obtained by cutting the dendrogram of the hierarchy).
\item A  random initialization:
\begin{enumerate}
\item $K$ variables are randomly selected among the $p$ variables as initial central synthetic variables (named centers hereafter).
\item An initial partition into $K$ clusters is built by allocating each variable to the cluster with the closest initial center: the similarity between a variable and an initial center is calculated using the function \code{mixedVarSim}.
\end{enumerate}
\end{enumerate} 
\item Repeat
\begin{enumerate}
\item A representation step: the quantitative central synthetic variable $\by_k$  of each cluster $C_k$ is calculated with PCAMIX as defined in section~\ref{sec:hom}.
\item An allocation step: a partition is constructed by assigning each variable to the closest cluster. The similarity  between a variable and the central synthetic quantitative variable of the corresponding cluster is calculated with   the function \code{mixedVarSim}: it is either a squared correlation (if the variable is quantitative) or a correlation ratio (if the variable is qualitative).
\end{enumerate}
\item Stop if there is no more changes in the partition or if a maximum number of iterations (fixed by the user) is  reached.
\end{enumerate}
This iterative procedure \code{kmeansvar} provides a partition $P_K$ into $K$ clusters   which maximizes  $\cal{H}$ but this optimum is local and depends on the initial partition. A solution to overcome this problem and to avoid the influence of the choice of an arbitrary initial partition is to consider multiple random initializations. In this case, steps 1(b), 2 and 3 are repeated, and we propose to retain as final  partition the one which provides the highest value of $\cal{H}$. 

\paragraph{Stability of partitions of variables.} This procedure evaluates the stability of the $p$ nested partitions of the dendrogram obtained with \code{hclustvar}. It works as follows:
\begin{enumerate}
\item $B$ boostrap samples of the $n$ observations are drawn and the corresponding $B$ dendrograms are obtained with the function \code{hclustvar}.
\item The partitions of these $B$  dendrograms are compared with the partitions of the initial hierarchy using the corrected Rand index. The Rand and the adjusted Rand indices are implemented in the function \code{Rand} (see \cite{huber85} for details on these indices).
\item The stability of a partition is evaluated by the mean of the $B$ adjusted Rand indices.
\end{enumerate}
The plot of this stability criterion according to the number of clusters can help the user in the choice of a sensible and suitable number of clusters. Note that an error message may appear with this function in some case of rare categories of qualitative variable. Indeed, if this rare category disappears in a bootstrap sample of observations, a column of identical values is then formed and the standardization of this variable is not possible in PCAMIX step.

%%%%%%%%%%%%%%%%%%%%%%%%%%%%%
\section{Illustration on simple examples} \label{sec:exple}
%%%%%%%%%%%%%%%%%%%%%%%%%%%%%

We illustrate our \proglang{R} package  \pkg{ClustOfVar} on two real datasets: the first one only concerns quantitative variables, the second one is a mixture of quantitative and qualitative variables.

%=====================================
\subsection{First example: Quantitative data} 
%=====================================

We use the dataset  \code{decathlon} which contains $n=41$ athletes described according to their performances in $p=10$ different sports of decathlon. 
\begin{CodeChunk}
\begin{CodeInput}
R> library("ClustOfVar")
R> data("decathlon")
R> head(decathlon[,1:4])
\end{CodeInput}
\begin{CodeOutput}
         100m Long.jump Shot.put High.jump
SEBRLE  11.04      7.58    14.83      2.07
CLAY    10.76      7.40    14.26      1.86
KARPOV  11.02      7.30    14.77      2.04
BERNARD 11.02      7.23    14.25      1.92
YURKOV  11.34      7.09    15.19      2.10
WARNERS 11.11      7.60    14.31      1.98
\end{CodeOutput}
\end{CodeChunk}
In order to have an idea of the links between these $10$ quantitative variables, we will construct a hierarchy with the function \code{hclustvar}.
\begin{CodeChunk}
\begin{CodeInput}
R> X.quanti <- decathlon[,1:10]
R> tree <- hclustvar(X.quanti)
R> plot(tree)
\end{CodeInput}
\end{CodeChunk}
\begin{figure}[h]
\centering
  \includegraphics[width=6cm,height=6cm]{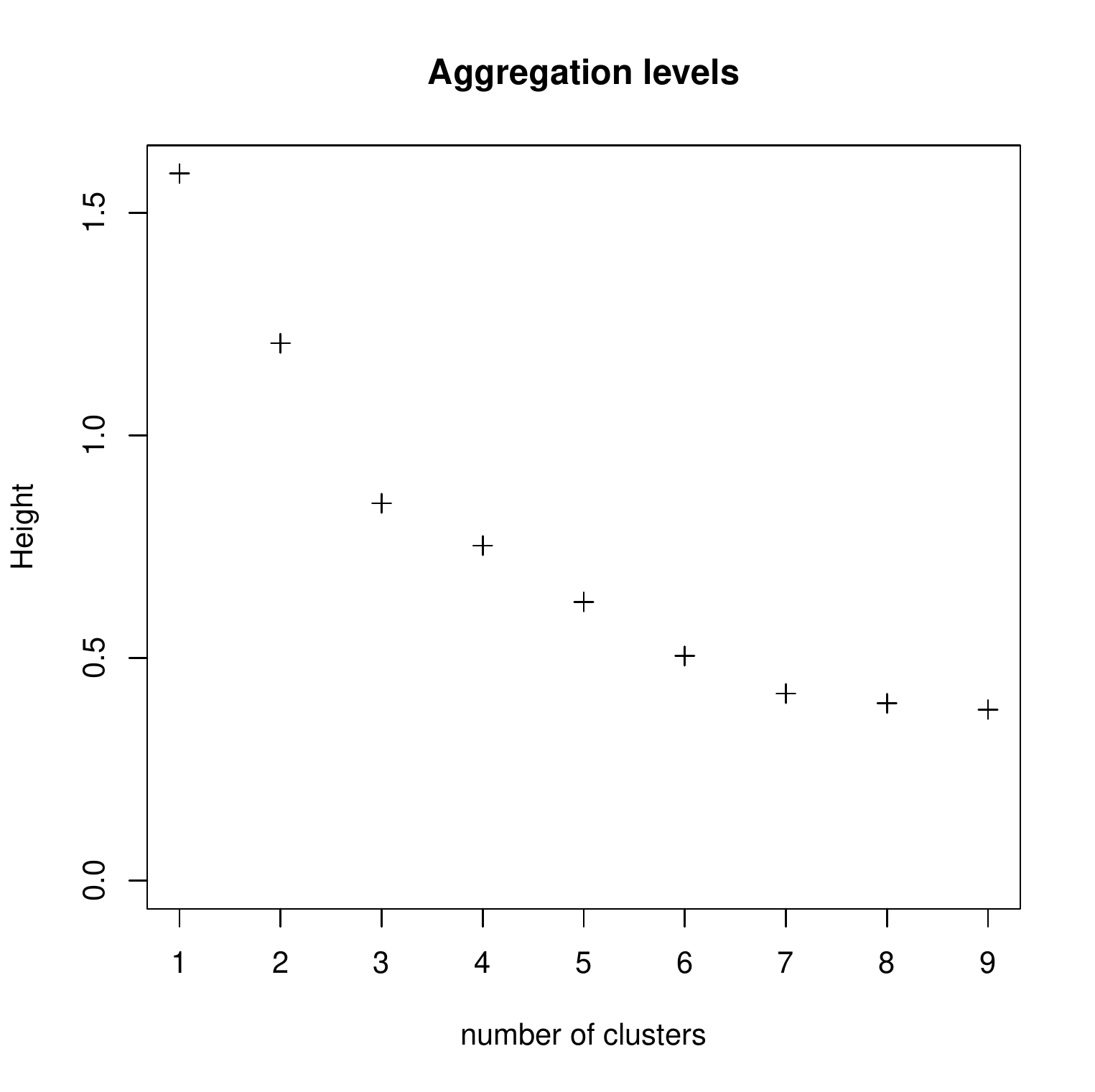}
 \includegraphics[width=6cm,height=6cm]{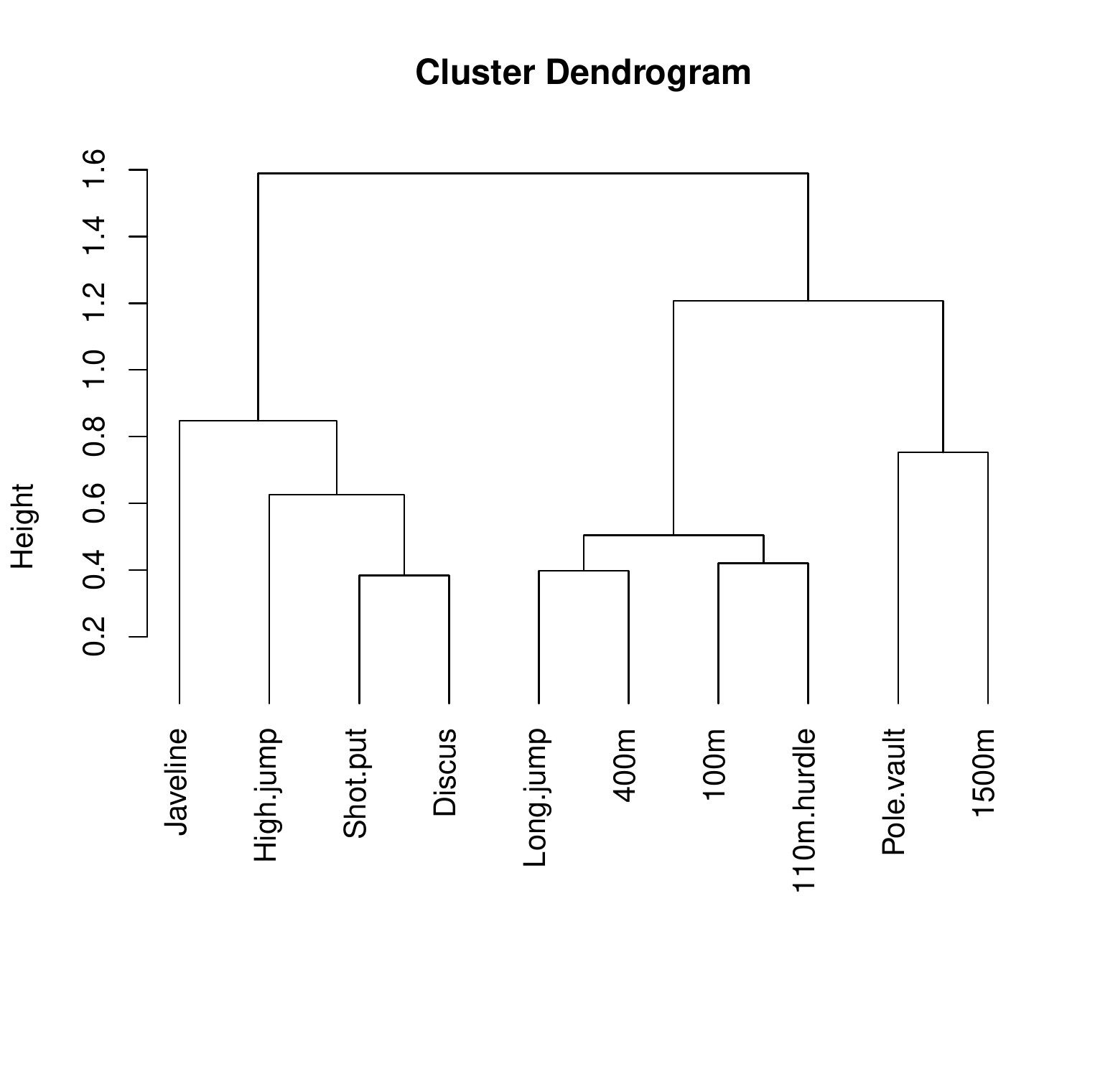}
\caption{Graphical output of the function \code{plot.hclustvar}.} \label{fig1}
\end{figure}
 In Figure~\ref{fig1},  the plot of the aggregation levels suggests to choose 3 or 5 clusters of variables.  
 The dendrogram, on the right hand side of this figure, shows the link between the variables in terms of $r^2$. For instance, the two variables ``discus'' and ``shot put''  are linked as well as the two variables  ``Long.jump'' and ``400m'', but the user must keep in mind that the dendrogram does not indicate the sign of these relationships: a careful study of these variables shows that ``discus'' and ``shot put'' are correlated whereas ``Long .jump'' and ``400m'' are anti-correlated.

\begin{figure}[h]
\centering
  \includegraphics[width=6cm,height=6cm]{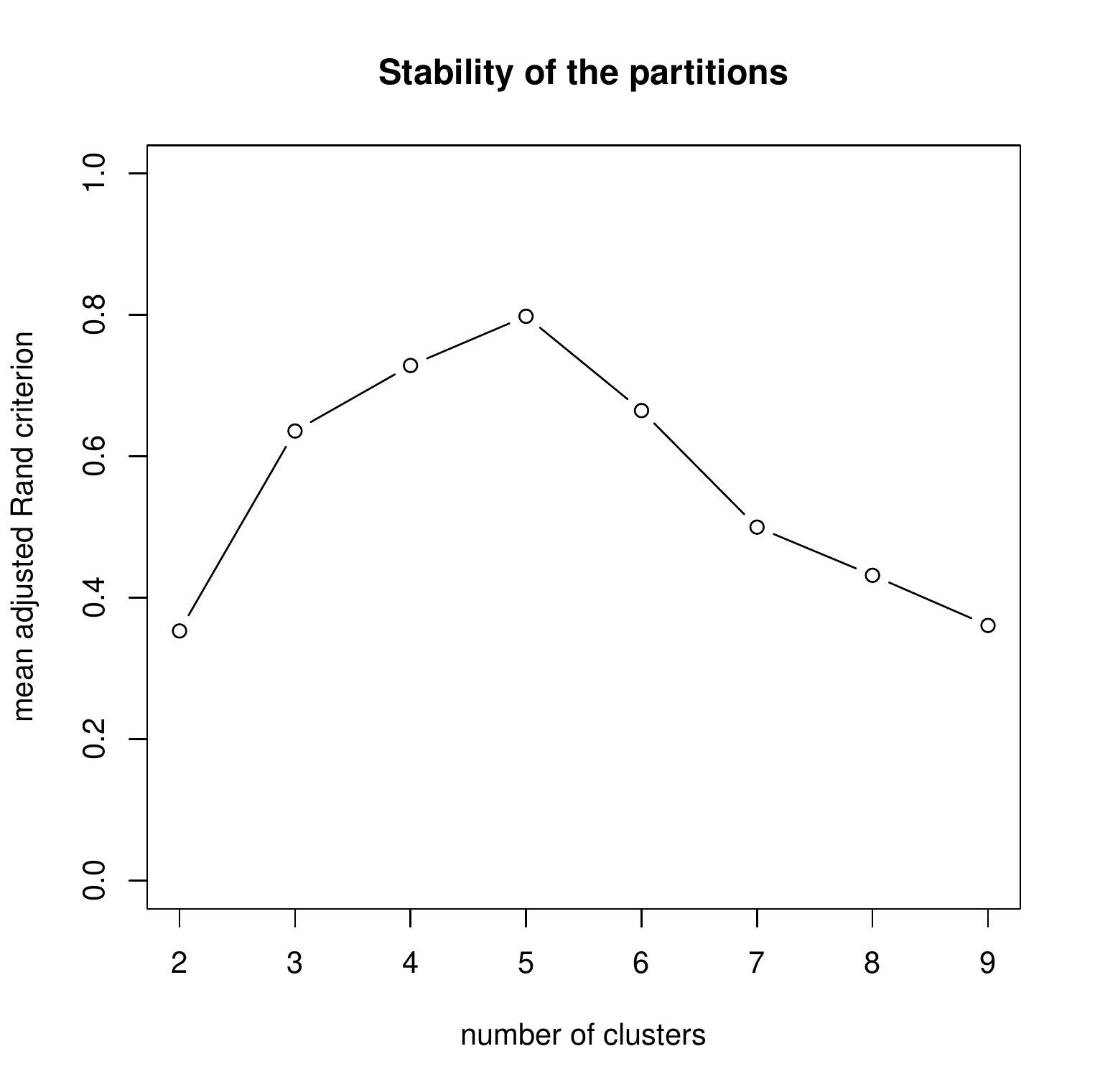}
 \includegraphics[width=6cm,height=6cm]{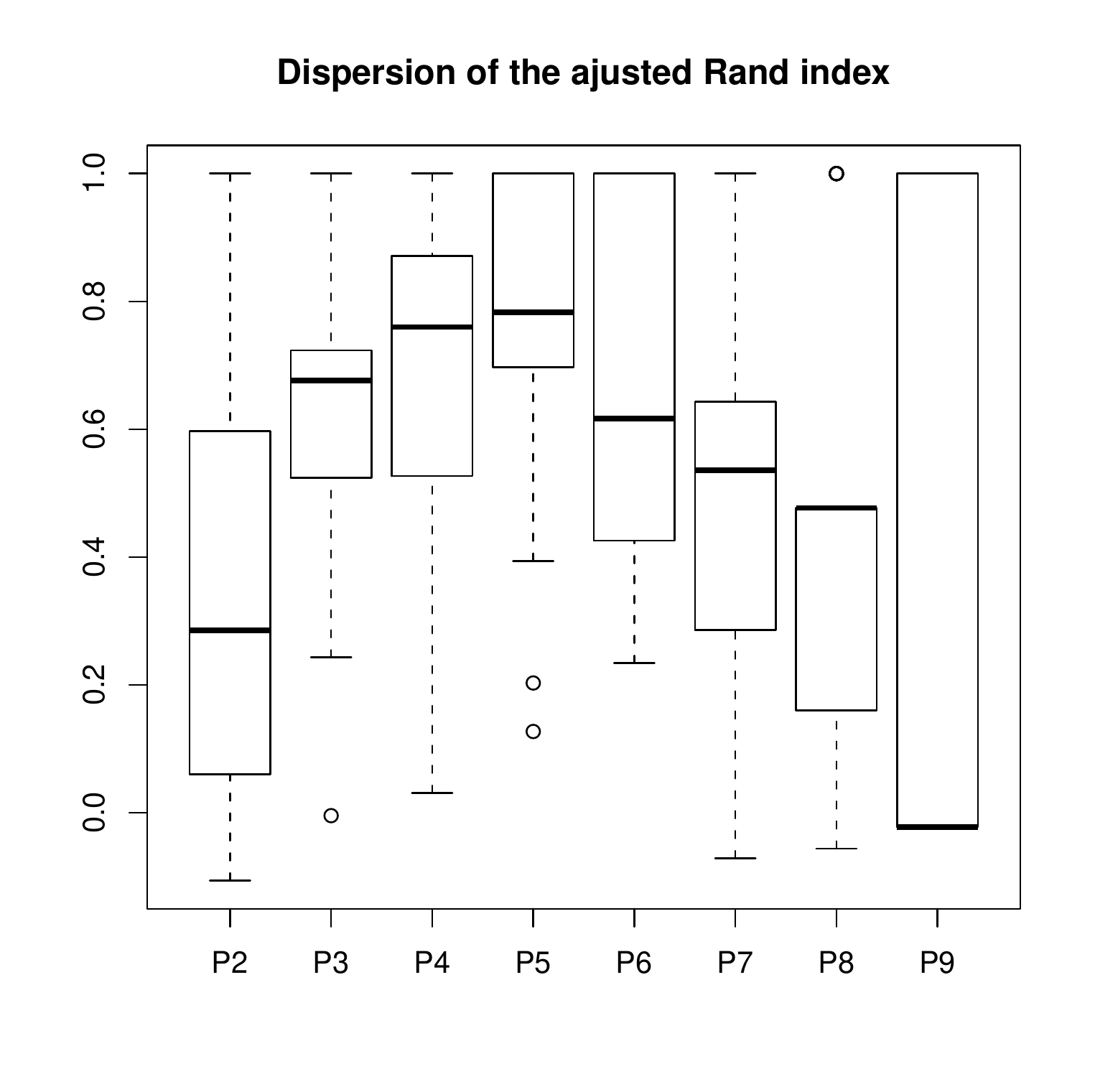}
\caption{Graphical output of the functions \code{stability} and \code{plot.clustab}.} \label{fig2}
\end{figure}

The user can use the \code{stability} function in order to have an idea of the stability of the partitions of the dendrogram represented in Figure~\ref{fig1}. 
\begin{CodeChunk}
\begin{CodeInput}
R> stab <- stability(tree,B=40)
R> plot(stab, main="Stability of the partitions")
R> stab$matCR
R> boxplot(stab$matCR, main="Dispersion of the ajusted Rand index")
\end{CodeInput}
\end{CodeChunk}
On the left of Figure~\ref{fig2}, the plot of the mean (over the $B=40$ bootstrap samples) of the adjusted Rand indices  is obtained with the function \code{plot.clustab}. It clearly suggests to choose 5 clusters. The boxplots on the right of Figure~\ref{fig2} show  the dispersion of these indices over the $B=40$ bootstrap replications for partition, and they suggest 3 or 5 clusters.

In the following we choose $K=3$ clusters because PCA applied to each of the 3 clusters gives each time only one eigenvalue greater than 1. The function \code{cutree} cuts the dendrogram of the hierarchy and gives a partition into $K=3$ clusters of the $p=10$ variables: 
\begin{CodeChunk}
\begin{CodeInput}
R> P3<-cutreevar(tree,3)
R> cluster <- P3$cluster
R> princomp(X.quanti[,which(cluster==1)],cor=TRUE)$sdev^2
R> princomp(X.quanti[,which(cluster==2)],cor=TRUE)$sdev^2
R> princomp(X.quanti[,which(cluster==3)],cor=TRUE)$sdev^2
\end{CodeInput}
\end{CodeChunk}

 The partition$P3$  is contained in an object  of class \code{clustvar}. Note that partitions obtained with the \code{kmeansvar} function are also objects  of class \code{clustvar}. The function \code{print.clustvar} gives a description of the values of  this object.
\begin{CodeChunk}
\begin{CodeInput}
R> P3<-cutreevar(tree,3,matsim=TRUE)
R> print(P3)
\end{CodeInput}
\begin{CodeOutput}
Call:
cutreevar(obj = tree, k = 3)

 name       description                        
 "$var"     "list of variables in each cluster"
 "$sim"     "similarity matrix in each cluster"
 "$cluster" "cluster memberships"              
 "$wss"     "within-cluster sum of squares"    
 "$E"       "gain in cohesion (in %)"          
 "$size"    "size of each cluster"             
 "$scores"  "score of each cluster"            
\end{CodeOutput}
\end{CodeChunk}

The value \code{$wss} is ${\cal H}(P_K)$ where the homogeneity function ${\cal H}$ was defined in (\ref{critP}). The gain in cohesion  \code{$E} is the percentage of homogeneity which is accounted by the partition  $P_K$. It is defined by:
\begin{equation}
E(P_K)=\frac{{\cal H}(P_K)-{\cal H}(P_1)}{p-{\cal H}(P_1)}. \label{gain}
\end{equation}
The value \code{$sim} provides the similarity matrices of the variables in each cluster (calculated with the function \code{mixedVarSim}). Note that it is time consuming to perform these similarity matrices when the number of variables is large. Thus they are not calculated by default: \code{matsim=TRUE} must be specified in the parameters of the function \code{hclustvar} (or \code{kmeansvar}) if the user wants this output.
We provide below the similarity matrix for the first cluster of this partition into 3 clusters.
\begin{CodeChunk}
\begin{CodeInput}
> round(P3$sim$cluster1,digit=2)
\end{CodeInput}
\begin{CodeOutput}
            100m Long.jump 400m 110m.hurdle
100m        1.00      0.36 0.27        0.34
Long.jump   0.36      1.00 0.36        0.26
400m        0.27      0.36 1.00        0.30
110m.hurdle 0.34      0.26 0.30        1.00
\end{CodeOutput}
\end{CodeChunk}

The value \code{$cluster} is a vector of integers indicating the cluster to which each variable is allocated.
\begin{CodeChunk}
\begin{CodeInput}
R> P3$cluster
\end{CodeInput}
\begin{CodeOutput}
       100m   Long.jump    Shot.put   High.jump        400m 110m.hurdle 
          1           1           2           2           1           1 
     Discus  Pole.vault    Javeline       1500m 
          2           3           2           3 
\end{CodeOutput}
\end{CodeChunk}
The value \code{$var} gives a description of each cluster of the partition.  More precisely it provides for each cluster the squared loadings on the first principal component of PCAMIX (which is the central synthetic variable of this cluster). For quantitative variables  (resp. qualitative), the squared loadings are squared correlations (resp. correlation ratio) with this central synthetic variable. For instance the squared correlation  between the variable ``100m'' and the central synthetic variable of  ``cluster1''  is 0.68.
\begin{CodeChunk}
\begin{CodeInput}
R> P3$var
\end{CodeInput}
\begin{CodeOutput}
$cluster1
            squared loading
100m              0.6822349
Long.jump         0.6873076
400m              0.6652279
110m.hurdle       0.6427661

$cluster2
          squared loading
Shot.put        0.7861012
High.jump       0.4991778
Discus          0.6023186
Javeline        0.2546550

$cluster3
           squared loading
Pole.vault       0.6237239
1500m            0.6237239
\end{CodeOutput}
\end{CodeChunk}

The value \code{$scores}  is the $n \times K$ matrix of the scores of the $n$ observations on the  first principal components of PCAMIX applied to the $K$ clusters: PCAMIX is applied 3 times here, one time in each cluster. Each column is then a synthetic variable of a cluster. The central synthetic variable of  ``cluster1'' for instance is  the first column of the $41 \times 3$ matrix above.  This column gives the scores of the 41 athletes on the first component of PCAMIX applied to the variables of ``cluster1'' (100m, Long.jump, 400m, 110m.hurdle).
\begin{CodeChunk}
\begin{CodeInput}
R> head(part_hier$scores)
\end{CodeInput}
\begin{CodeOutput}
          cluster1   cluster2   cluster3
SEBRLE   0.2640687 -1.0353928 -1.4405915
CLAY     1.3816943 -0.3454687 -1.7840860
KARPOV   1.1098485 -0.7209119 -1.7043603
BERNARD -0.1949061  0.7082857 -1.5017373
YURKOV  -2.0319539 -1.8850107  0.2702640
WARNERS  1.1385110  1.0929346 -0.3490226
\end{CodeOutput}
\end{CodeChunk}
 Note that this $41 \times 3$ matrix of the scores of the  $41$ athletes in each cluster of variables is of course different from the  $41 \times 3$ matrix of the scores  of the athletes on the  first 3 principal components of PCAMIX (here PCA) applied to the initial dataset. The 3 synthetic variables for instance can be correlated whereas the first 3 principal components of PCAMIX are not correlated by construction. But the matrix of the synthetic variables in  \code{$scores} can be used as the matrix of the principal components of PCAMIX for dimension reduction purpose.

%==========================================================================
\subsection{Second example: A mixture of quantitative and qualitative data} 
%==========================================================================

\begin{figure}[htb]
\centering
 \includegraphics{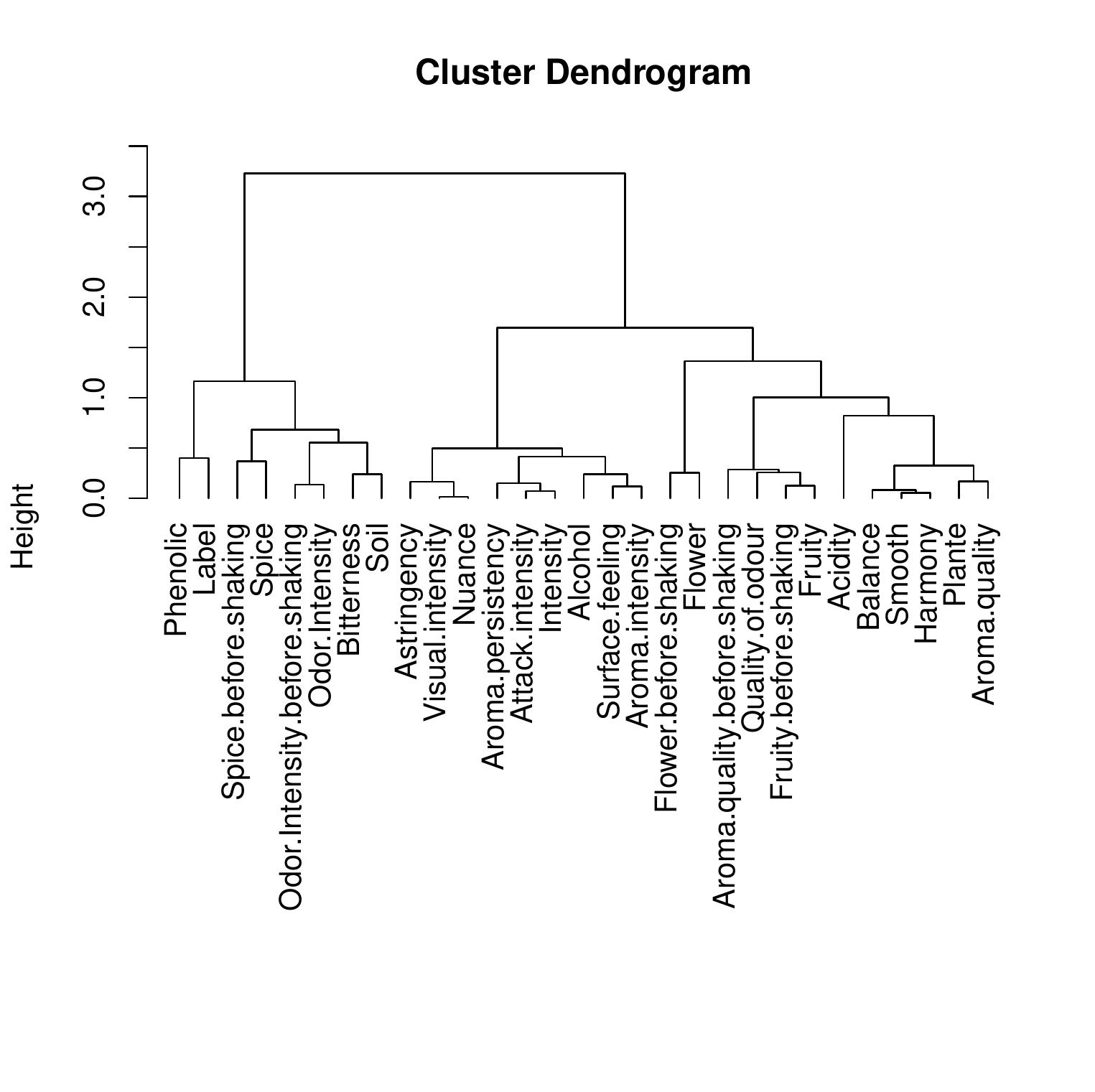}
\caption{Dendrogram of the hierarchy of the 31 variables of the wine dataset.} \label{fig3}
\end{figure}

We use the dataset  \code{wine} which contains $n=21$ french wines described by $p=31$ variables. The first two  variables ``Label'' and ``Soil'' are qualitative with respectively 3 and 4 categories. The other 29 variables are quantitative. 
\begin{CodeChunk}
\begin{CodeInput}
R> data("wine")
R> head(wine[,c(1:4)])
\end{CodeInput}
\begin{CodeOutput}
          Label      Soil Odor.Intensity Aroma.quality
2EL      Saumur      Env1          3.074         3.000
1CHA     Saumur      Env1          2.964         2.821
1FON Bourgueuil      Env1          2.857         2.929
1VAU     Chinon      Env2          2.808         2.593
1DAM     Saumur Reference          3.607         3.429
2BOU Bourgueuil Reference          2.857         3.111
\end{CodeOutput}
\end{CodeChunk}
 In order to have an idea of the links between these $31$ quantitative and qualitative variables, we construct a hierarchy using the function \code{hclustvar}.
\begin{CodeChunk}
\begin{CodeInput}
R> X.quanti <- wine[,c(3:29)] 
R> X.quali <- wine[,c(1,2)] 
R> tree <- hclustvar(X.quanti,X.quali)
R> plot(tree)
\end{CodeInput}
\end{CodeChunk}
In Figure~\ref{fig3}, we plot the dendrogram. It shows  for instance  that the qualitative variable ``label'' is linked (in term of correlation ratio) with the quantitative variable ``Phenolic''. 
The user chooses according to this dendrogram to cut this dendrogram into $K=6$ clusters: 
\begin{CodeChunk}
\begin{CodeInput}
R> part_hier<-cutreevar(tree,6)
R> part_hier$var$"cluster1"
\end{CodeInput}
\begin{CodeOutput}
                     squared loading
Odor.Intensity             0.7617528
Spice.before.shaking       0.6160243
Odor.Intensity.1           0.6663325
Spice                      0.5357837
Bitterness                 0.6620632
Soil                       0.7768805
\end{CodeOutput}
\end{CodeChunk}
A close reading of the output for  ``cluster1'' shows that the correlation ratio between the qualitative variable ``Soil'' and the synthetic variable of the cluster is about 0.78. The squared correlation between ``Odor.Intensity'' and the synthetic variable of the cluster is 0.76. 

The  central synthetic variables of the 6 clusters are in \code{part_hier$scores}. This $21 \times 6$ quantitative matrix  can replace the original $21 \times 31$ data matrix mixing qualitative and quantitative variables. This matrix of the synthetic variables  can then be used  for recoding a mixed data matrix (or a qualitative data matrix) into a quantitative data matrix, as is usually done with the matrix of the principal components of PCAMIX. 

The function \code{kmeansvar}  can also provide a partition into $K=6$ clusters of the $31$ variables. 
\begin{CodeChunk}
\begin{CodeInput}
R> part_km<-kmeansvar(X.quanti,X.quali,init=6,nstart=10)
\end{CodeInput}

The gain in cohesion of the partition  in (\ref{gain}) obtained with the k-means type partitioning algorithm and 10 random initializations is smaller than that of the partition obtained with the hierarchical clustering algorithm (51.02 versus 56.84): 
\begin{CodeOutput}
R> part_km$E
 [1] 51.02414
R> part_hier$E
 [1] 56.84082
\end{CodeOutput}
\end{CodeChunk}

In practice, simulations and real datasets showed that the quality of the partitions obtained with \code{hclustvar} seems to be better than that obtained with \code{kmeansvar}. But for large datasets (with a large number of variables), the  function  \code{hclustvar} meets problems of computation time. In this case, the  function \code{kmeansvar} will be faster.

%%%%%%%%%%%%%%%%%%%%%%%%%%%%%
\section{Concluding remarks} \label{sec:conclu}
%%%%%%%%%%%%%%%%%%%%%%%%%%%%%

The \proglang{R} package  \pkg{ClustOfVar} proposes hierarchical and k-means type algorithms for the clustering of variables of any type (quantitative and/or qualitative).  

This package proposes  useful tools to visualize the links between the variables and the redundancy in a data set. It is also an alternative to principal component analysis methods for  dimension reduction and  for recoding qualitative or mixed data matrices into quantitative data matrix. The main difference between PCA and the approach of clustering of variables presented in this paper, is that the synthetic variables of the clusters can be correlated whereas the principal components are not correlated by construction. 

The package \pkg{ClustOfVar} is not performing well with datasets having very large number of variables:  the computational time becomes relatively long.  A future work is to propose a new version of the package with versions of the functions \code{hclustvar}, \code{kmeansvar} and \code{stability}  developed for parallel computing.

We mention that the package  \pkg{ClustOfVar} can deal with missing data. However let us note that the imputation method used in the code is  simple and may not perform well when the proportion of missing data is too large. In that case, one of the numerous \proglang{R} packages  devoted to missing data imputation should be used prior to \pkg{ClustOfVar}.  

\bibliography{JSS_ClustOfVar}
\end{document}